\documentclass[12pt]{article}
\usepackage{epsfig}
\usepackage{amssymb}
\usepackage{amsmath}
\usepackage{amsfonts}

\newcommand{\R}{\mathbb{R}}
\newcommand{\C}{\mathbb{C}}
\newcommand{\Z}{\mathbb{Z}}

\newcommand{\be}{\begin{equation}}
\newcommand{\ee}{\end{equation}}
\newcommand{\bea}{\begin{eqnarray}}
\newcommand{\eea}{\end{eqnarray}}
\newcommand{\nn}{\nonumber}
\newcommand{\kt}{\rangle}
\newcommand{\br}{\langle}

\newcommand{\ed}{\end{document}}

\begin{document}

\title{Differential Realization of Pseudo-Hermiticity:\\
A quantum mechanical analog of\\ Einstein's field equation}

\author{\\
Ali Mostafazadeh
\\
\\
Department of Mathematics, Ko\c{c} University,\\
34450 Sariyer, Istanbul, Turkey\\ amostafazadeh@ku.edu.tr}
\date{ }
\maketitle

\begin{abstract}
For a given pseudo-Hermitian Hamiltonian of the standard form:
$H=p^2/2m+v(x)$, we reduce the problem of finding the most general
(pseudo-)metric operator $\eta$ satisfying $H^\dagger=\eta
H\eta^{-1}$ to the solution of a differential equation. If the
configuration space is $\R$, this is a Klein-Gordon equation with
a nonconstant mass term. We obtain a general series solution of
this equation that involves a pair of arbitrary functions. These
characterize the arbitrariness in the choice of $\eta$. We apply
our general results to calculate $\eta$ for the ${\cal
PT}$-symmetric square well, an imaginary scattering potential, and
a class of imaginary delta-function potentials. For the first two
systems, our method reproduces the known results in a
straightforward and extremely efficient manner. For all these
systems we obtain the most general $\eta$ up to second order terms
in the coupling constants.

\vspace{5mm}

\noindent PACS number: 03.65.-w\vspace{2mm}

\noindent Keywords: metric operator, pseudo-Hermitian,
quasi-Hermitian, ${\cal PT}$-symmetry, complex potential,
delta-function potential

\end{abstract}




\newpage

\section{Introduction}

The key aspect of the General Theory of Relativity (GR) that
distinguishes it from other well-established physical theories is
that in GR the very geometry of the spacetime, which is the arena
of physical reality in classical physics, is itself a dynamical
quantity. In contrast in Quantum Mechanics (QM) the geometry of
the Hilbert space, which plays a similar role as the spacetime
does in classical physics, is an absolute entity. This is to some
extent dictated by the well-known mathematical fact that all
separable Hilbert spaces are unitary-equivalent. For the past 75
years or so, this equivalence has been used to justify the
absolutism associated with the convention of fixing the (inner
product of the) Hilbert space from the outset. This is actually
quite surprising, for the existence of an equivalence relation in
a theory is clearly an evidence of the presence of a freedom in
its formulation. In the case of QM, this is the freedom to choose
the inner product of the Hilbert space, a freedom that has been
left unused till recently, \cite{geyer-92,pla-04}.

In \cite{pla-04}, we have investigated the consequences of
promoting the inner product of the Hilbert space into a degree of
freedom. This revealed certain similarities between QM and GR and
led to some interesting observations such as a direct link between
geometric phases and the geometry of the Hilbert space and a new
root to a certain nonlinear generalization of QM. In the present
paper, we derive and examine a differential equation that includes
among its solutions all possible choices of the inner product for
a given physical system. This is the quantum mechanical analogue
of Einstein's field equation.\footnote{The same way Einstein's
equation does not generally restrict the metric tensor to have a
particular signature, the above-mentioned equation does not
restrict its solutions to correspond to positive-definite metric
operators.} For a system having $\R$ as its configuration space we
obtain a series solution of this equation that involves two
functional degrees of freedom. These signify the arbitrariness in
the choice of the (pseudo-)metric operator. Our approach allows a
more direct way of addressing some of the basic practical problems
arising in the application of quasi- and pseudo-Hermitian quantum
mechanics,
\cite{geyer-92,cjp-2003a,jpa-2004b,jpa-2005b,jones-05,jmp-2005,bagchi-q-r,p67,scholtz-geyer,bender-tan,jones-06}.
In particular it provides an extremely powerful technical tool for
the perturbative calculation of the (pseudo-)metric operators for
various toy models.

\section{Differential Representation of Pseudo-Hermiticity}

Consider a physical system described by a separable Hilbert space
${\cal H}$ and a pseudo-Hermitian Hamiltonian operator $H:{\cal
H}\to{\cal H}$. Let ${\cal M}$ denote the set of all linear
invertible Hermitian operators $\eta:{\cal H}\to{\cal H}$, then by
definition \cite{p1} the pseudo-Hermiticity of $H$ means that
${\cal M}_{_H}:=\{\eta\in{\cal M}| H^\dagger=\eta\,H\eta^{-1}\}$
is a nonempty subset of ${\cal M}$. The elements of ${\cal M}$ are
called \emph{pseudo-metric operators}, for they may be used to
define a pseudo-inner product (a nondegenerate sesquilinear form
\cite{kato}) $\br\cdot|\cdot\kt_{_\eta}:=\br\cdot|\eta\cdot\kt$ on
${\cal H}$, where $\br\cdot|\cdot\kt$ denotes the defining inner
product of ${\cal H}$.\footnote{Strictly speaking,
$\br\cdot|\cdot\kt_{_\eta}$ is a nondegenerate sesquilinear form
defined on the domain of $\eta$.}

The Hamiltonian $H$ is Hermitian with respect to
$\br\cdot|\cdot\kt_{_\eta}$ for all $\eta\in{\cal M}_{_H}$,
\cite{pauli-1943,p1}. If ${\cal M}_{_H}$ includes a
positive-definite element, i.e., a \emph{metric operator},
$\eta_+$, then $H$ is Hermitian with respect to the
positive-definite inner product $\br\cdot|\cdot\kt_{_{\eta_+}}$.
This implies that $H$ is diagonalizable and has a real spectrum.
The converse of this statement holds true at least for the case
that the spectrum of $H$ is discrete, i.e., if $H$ is
diagonalizable and has a real spectrum then ${\cal M}_{_H}$
includes a positive-definite element and equivalently $H$ is
Hermitian with respect to a positive-definite inner product,
\cite{p2,p3}. Furthermore, in this case one can show that $H$ is
necessarily quasi-Hermitian, i.e., it may be mapped to a Hermitian
Hamiltonian $h:{\cal H}\to{\cal H}$ via a similarity
transformation, $H=\rho^{-1}h\,\rho$, \cite{p2,p3}. This and only
this class of Hamiltonians are capable of supporting a unitary
time-evolution in an associated \emph{physical Hilbert space}. The
latter is defined by endowing ${\cal H}$ with the inner product
$\br\cdot|\cdot\kt_{_{\eta_+}}$ and will be denoted by ${\cal
H}_{_{\eta_+}}$, \cite{critique,jpa-2004b,jpa-2005b,jmp-2005,p67}.

An important fact about this construction is that $\eta_+$ is not
unique. Different choices for $\eta_+$ yield kinematically
distinct quantum systems that nevertheless share the same
dynamical structure. The quantum mechanical analogue of the
principle of general covariance of GR is the physical (unitary-)
equivalence of quantum systems $({\cal H}_{_{\eta_+}},H)$,
\cite{pla-04}. The metric operators $\eta_+$ and more generally
pseudo-metric operators $\eta$ are linked to and consequently
determined by the Hamiltonian $H$ via the pseudo-Hermiticity
condition
    \be
    H^\dagger=\eta\,H\,\eta^{-1}.
    \label{ph}
    \ee
The same way Einstein's field equation links the metric tensor to
the energy-momentum tensor, (\ref{ph}) links the pseudo-metric
operator to the Hamiltonian. The resemblance may be made more
pronounced for a Hamiltonian of the standard form,
    \be
    H=\frac{\vec p^2}{2m}+v(\vec x),
    \label{H=}
    \ee
that acts in $L^2(\R^n)$. Applying both sides of (\ref{ph}) on
$\eta$, substituting (\ref{H=}), and representing the resulting
equation in the $\vec x$-basis, we find
    \be
    \left(-\nabla_x^2+\nabla_y^2+\frac{2m}{\hbar^2}[
    v(\vec x)^*-v(\vec y)]\right)\eta(\vec x,\vec y)=0,
    \label{pde}
    \ee
where $\nabla_u^2:=\sum_{j=1}^n \partial^2/\partial u_j^2$ for
$u=x,y$ and $\eta(\vec x,\vec y):=\br\vec x|\eta|\vec y\kt$. For
$n=1$, this is a Klein-Gordon equation with a variable mass term,
    \be
    \left[-\partial_x^2+\partial_y^2+\mu^2(x,y)\right]\eta(x,y)=0,
    ~~~~~~~~~~
    \mu^2(x,y):=\frac{2m}{\hbar^2}[
    v(x)^*-v(y)].
    \label{pde-1}
    \ee

According to (\ref{pde}) if $\eta(\vec x,\vec y)$ is a solution,
then so is $\eta(\vec y,\vec x)^*$. The pseudo-metric operators
$\eta\in{\cal M}_{_H}$ correspond to solutions that satisfy
    \be
    \eta(\vec x,\vec y)^*=\eta(\vec y,\vec x).
    \label{hermitian}
    \ee
Note that even for non-pseudo-Hermitian Hamiltonians of the form
(\ref{H=}), Eq.~(\ref{pde}) admit solutions. However, these
solutions fail to satisfy either the Hermiticity requirement
(\ref{hermitian}) or the invertibility condition:
    \be
    \int_{\R^n} d^n\vec y~\eta(\vec x,\vec y)\,\psi(\vec y)=0
    ~~{\rm implies}~~\psi=0.
    \label{invert}
    \ee

If ${\cal M}_{_H}$ happens to include positive-definite elements
$\eta_+$, then these elements correspond to the solutions
$\eta_+(\vec x,\vec y)$ of (\ref{pde}) that in addition to
(\ref{hermitian}) satisfy
    \be
    \int_{\R^n} d^n\vec x~\int_{\R^n} d^n\vec y~\psi(\vec x)^*\,
    \eta_+(\vec x,\vec y)\,\psi(\vec y)>0~~{\rm for}~~\psi\neq 0.
    \label{positive-def}
    \ee
The fact that for a diagonalizable Hamiltonian with a real and
discrete spectrum such solutions exist is a consequence of the
spectral theorems given in \cite{p2,p3}. Indeed if
$\{\psi_n,\phi_n\}$ is a biorthonormal system associated with $H$,
i.e., $H\psi_n=E_n\psi_n$, $H^\dagger \phi_n=E_n\phi_n$,
$\br\psi_n|\phi_m\kt=\delta_{mn}$, then
    \be
    \eta_+(\vec x,\vec y)=\sum_n\phi_n(\vec x)\phi_n(\vec y)^*
    \label{spec}
    \ee
is a solution of (\ref{pde}) that satisfies both (\ref{hermitian})
and (\ref{positive-def}). As shown in \cite{npb-2002,jmp-2003}, in
this case the most general (positive-definite) metric operator has
the form $A^\dagger\eta_+A$ where $A$ is invertible and commutes
with $H$. The latter corresponds to a solution of (\ref{pde}) that
is of the form
    \be
    \eta'_+(\vec x,\vec y)=\sum_n \int_{\R^n} d^n\vec u~\int_{\R^n}
    d^n\vec v~ A(\vec u,\vec x)^* \phi_n(\vec u)\phi_n(\vec v)^*
    A(\vec v,\vec y),
    \label{spec2}
    \ee
where $A(\vec x,\vec y)$ satisfies
    \be
    \left(-\nabla_x^2+\nabla_y^2+\frac{2m}{\hbar^2}[
    v(\vec x)-v(\vec y)]\right)A(\vec x,\vec y)=0,
    \label{pde-A}
    \ee
and
    \be
    \int_{\R^n} d^n\vec y~A(\vec x,\vec y)\,\psi(\vec y)=0
    ~~{\rm implies}~~\psi=0.
    \label{invert-A}
    \ee

For a real-valued potential $v$, $\delta(\vec x-\vec y)$ is a
solution of (\ref{pde}). It corresponds to the choice $\eta=I$,
where $I$ is the identity operator acting in ${\cal H}=L^2(\R)$.
This is consistent with the Hermiticity of $H$. For a ${\cal
PT}$-symmetric potential $v$ that satisfies $v(-\vec x)=v(\vec
x)^*$, $\delta(\vec x+\vec y)$ is a solution of (\ref{pde}). This
is a manifestation of ${\cal P}$-pseudo-Hermiticity of the
Hamiltonian~\cite{p1}, for $\br\vec x|{\cal P}|\vec
y\kt=\delta(\vec x+\vec y)$.

For $n>1$, (\ref{pde}) is an ultra-hyperbolic equation with quite
peculiar properties \cite{ultra}. We will therefore, focus our
attention on the case $n=1$. Our main purpose is to obtain the
general solution of (\ref{pde}) without having to resort to its
spectral decomposition (\ref{spec}). This is mainly because of the
difficulties with summing the series in (\ref{spec}) or evaluating
the integrals that replace the latter whenever the spectrum
becomes continuous \cite{jmp-2005}.

In \cite{p67}, we have pursued a similar approach to construct the
most general $\eta_+$ for the imaginary cubic potential
$v=i\epsilon x^3$ in some low orders of perturbation theory. The
approach of \cite{p67} applies to any (preferably imaginary)
potential with a real spectrum. It yields an infinite system of
iteratively decoupled partial differential equations whose
solution provides the contributions to $\eta_+$ in various orders
of the perturbation theory. Although these equations have the same
structure, at each order one must compute their non-homogeneous
term and solve them separately. In contrast, in the present paper,
we obtain a single differential equation satisfied by $\eta$,
namely (\ref{pde-1}), that applies for an arbitrary potential
$v(x)$ rendering the Hamiltonian pseudo-Hermitian. An important
advantage of the approach of the present paper over that of
\cite{p67} is that in view of the simple structure of
(\ref{pde-1}), we are able to offer (as discussed in Section~3) a
general scheme for constructing a series solution of this
equation. This solution involves two arbitrary functions that
provide an explicit characterization of the arbitrariness in the
choice of $\eta$.

Another recent application of the powerful machinery of
differential equations to compute (pseudo-)metric operators is due
to Scholtz and Geyer \cite{scholtz-geyer}. These authors obtain a
phase-space representation of the (pseudo-)metric operators
$\eta$. They use the Moyal product techniques to deal with the
difficult factor-ordering problems that arise in this
representation. The following are the main differences between the
method of \cite{scholtz-geyer} and the one presented in the
present paper.
    \begin{itemize}
    \item The method of \cite{scholtz-geyer} leads to an
equation for $\eta$ that is a differential equation
\cite{bender-dune} provided that $v(x)$ is a polynomial potential.
Even for a polynomial potential the general character and in
particular the order of this differential equation depends on the
structure of $v(x)$ and its degree. In contrast, in the present
paper we offer a universal differential equation, namely
(\ref{pde-1}), that applies for polynomial as well as
non-polynomial potentials, and has the same simple structure for
all potentials. It is this appealing property that allows us to
treat the well-known toy models of Section~4. The application of
the method of \cite{scholtz-geyer} to these models yields
pseudo-differential equations (differential equations of infinite
order) whose solution is extremely difficult if not impossible.
    \item Suppose $v$ is a polynomial potential, so that the
method of \cite{scholtz-geyer} yields a differential equation, and
suppose that one is able to solve this equation. Then one obtains
an explicit expression for $\eta$ in terms of the operators $x$
and $p$ which involves a number of arbitrary functions. The
condition that $\eta$ be Hermitian must be imposed to fix some of
these functions. This is done by adopting a set of appropriate
boundary conditions \cite{scholtz-geyer}.\footnote{The author is
unaware of a systematic method of selecting the boundary
conditions that achieve this purpose.} In contrast, our method
yields an expression for $\eta(x,y)$ that satisfies the
Hermiticity condition  $\eta(x,y)^*=\eta(y,x)$ manifestly and
specifies a unique Hermitian $\eta$ according to
    \[(\eta\psi)(x)=\int_{\R}dy~\eta(x,y)\psi(y).\]
It achieves this without making use of the Moyal product or having
to select certain boundary conditions that ensures the Hermiticity
of $\eta$. Its successful application, however, does not yield an
explicit expression for $\eta$ in terms of $x$ and $p$. As
explained in \cite{p67}, the latter may be obtained by Fourier
transforming $\eta(x,y)=\br x|\eta|y\kt$ over $y$ to obtain $\br
x|\eta|p\kt$ and arranging the terms in the expression for
$\sqrt{2\pi\hbar}~e^{-ixp/\hbar}\br x|\eta|p\kt$ in such a way
that $x$'s are placed to the left of $p$'s. This is how the issue
of ordering of factors is addressed in this construction.
    \end{itemize}
A common feature of both methods is that solving the associated
differential equations yields generally non-positive-definite
pseudo-metric operators. The (positive-definite) metric operators
$\eta_+$, if they exist, correspond to certain special solutions
that are to be identified using different means.\footnote{The
construction of $\eta$ given here may be supplemented with the
procedure proposed in \cite{scholtz-geyer} for selecting the
positive-definite metric operators $\eta_+$ among $\eta$'s. This
is expected to be a difficult task in practice, and we will not
pursue it here. We suffice to point out that given $\eta(x,y)$ we
can obtain an expression for $\eta$ in terms of $x$ and $p$ as
outlined in \cite{p67}. This allows for making direct contact with
the approaches of \cite{scholtz-geyer} and \cite{bender-dune}.}

\section{Series Expansion for $\eta(x,y)$}

We begin our analysis by expressing (\ref{pde-1}) in the form
    \be
    \left(-\partial_x^2+\partial_y^2\right)\eta(x,y)=f(x,y)-f(y,x)^*
    \label{e1n}
    \ee
where
    \be
    f(x,y):=\frac{2m}{\hbar^2}\,v(y)\eta(x,y).
    \label{e2n}
    \ee
We note that for a Hermitian $\eta$, (\ref{e1n}) is equivalent to
    \bea
    &&\eta(x,y)=\chi(x,y)+\chi(y,x)^*,
    \label{e3n}\\
    &&\left(-\partial_x^2+\partial_y^2\right)\chi(x,y)=f(x,y).
    \label{e4n}
    \eea
Next, we recall that the general solution of the  wave equation
$\left(-\partial_x^2+\partial_y^2\right)u(x,y)=0$ is given by
    \be
    u(x,y)=u_+(x-y)+u_-(x+y),
    \label{homog-wave}
    \ee
where $u_\pm:\R\to\C$ are a pair of arbitrary twice-differentiable
functions (or distributions). Consequently, the general solution
of (\ref{e1n}) has the form
    \be
    \eta(x,y)=u_+(x-y)+u_-(x+y)+\chi_p(x,y)+\chi_p(y,x)^*,
    \label{gen-eta}
    \ee
where $u_\pm$ satisfy $u_\pm(x)^*=u_\pm(\mp x)$ and $\chi_p(x,y)$
is a particular solution of (\ref{e4n}). The latter is a
non-homogeneous wave equation in 1+1 dimensions. It admits a
particular solution which in view of (\ref{e2n}) takes the form
    \be
    \chi_p(x,y)=\frac{m}{\hbar^2}\,\int^ydr \int_{x-y+r}^{x+y-r}
    ds~v(r)\,\eta(s,r).
    \label{e5n}
    \ee

Combining (\ref{homog-wave}) -- (\ref{e5n}), we find
    \be
    \eta(x,y)=u(x,y)+{\cal K}\eta(x,y),
    \label{int-eq}
    \ee
where ${\cal K}$ is the integral operator defined by
    \bea
    {\cal K}\eta(x,y)&:=&\frac{m}{\hbar^2}\,\left[
    \int^ydr \int_{x-y+r}^{x+y-r}
    ds~v(r)\,\eta(s,r)+
    \int^xdr \int_{-x+y+r}^{x+y-r}
    ds~v^*(r)\,\eta(s,r)^*\right]\nn\\
    &=&\frac{m}{\hbar^2}\,\left[
    \int^ydr \int_{x-y+r}^{x+y-r}
    ds~v(r)\,\eta(s,r)+
    \int^xds \int_{-x+y+s}^{x+y-s}
    dr~v(s)^*\,\eta(s,r)\right].
    \label{K=}
    \eea
In view of the analogy with the derivation of the
Lippmann-Schwinger equation \cite{bohm-qm}, it is not difficult to
see that (\ref{int-eq}) admits the following general series
solution
    \be
    \eta(x,y)=[I-{\cal K}]^{-1}u(x,y)=\sum_{\ell=0}^\infty
    {\cal K}^\ell u(x,y).
    \label{formal}
    \ee
Clearly, $\eta$ is determined in terms of the arbitrary functions
$u_\pm$.

For $v=0$, i.e., a free particle, $\eta(x,y)=u(x,y)$. As shown in
\cite{p67}, this is equivalent to
    \be
    \eta=L(p)+K(p){\cal P},
    \label{eta-zero}
    \ee
where $L(p)^\dagger=L(p)$ and $K(p)^\dagger={\cal P}K(p){\cal P}$,
equivalently $L$ and $K$ are respectively real-valued and ${\cal
PT}$-invariant\footnote{This means $K(r)^*=K(-r)$ for all
$r\in\R$.} functions.\footnote{They may be further restricted to
constants if one postulates the nonexistence of a hidden length
scale for the problem. See \cite{p67} for details.} They are
related to the Fourier transform\footnote{In our convention, the
Fourier transform of a function $\varphi$ is given by $\tilde
\varphi(k):= (2\pi)^{-1/2}\int_{-\infty}^\infty dx\,
e^{-ikx}\varphi(x)$.\label{foot}} $\tilde u_\pm$ of $u_\pm$
according to
    \be
    L(p)=\sqrt{2\pi}\;\tilde u_+(\mbox{$\frac{p}{\hbar}$}),~~~~
    K(p)=\sqrt{2\pi}\;\tilde u_-(\mbox{$-\frac{p}{\hbar}$}).
    \label{LK}
    \ee

For a real-valued potential the ordinary choice for the metric
operator that yields the $L^2$-inner product, i.e., $\eta=I$,
corresponds to setting
    \be
    u(x,y)=
    \delta(x-y)-\frac{m}{\hbar^2}\int^{\frac{x+y}{2}}dr~v(r).
    \label{u-for-L2}
    \ee
To see this, we first calculate ${\cal K}\delta(x-y)$ for an
arbitrary (possibly complex-valued) potential $v$. Using the
well-known properties of the step function:
    \be
    \theta(x):=\left\{\begin{array}{ccc}
    0&{\rm for}& x<0\\
    \frac{1}{2}&{\rm for}& x=0\\
    1&{\rm for}& x>0,\end{array}\right.
    \label{theta}
    \ee
we then find
    \be
    {\cal K}\delta(x-y)=\frac{m}{\hbar^2}\left(
    \int^{\frac{x+y}{2}}dr~\Re[v(r)]+i\,{\rm sign}(y-x)
    \int^{\frac{x+y}{2}}dr~\Im[v(r)]\right),
    \label{k-for-delta}
    \ee
where $\Re[v]$ and $\Im[v]$ respectively stand for the real and
imaginary parts of $v$, and
    \[{\rm sign}(x):=\theta(x)-\theta(-x)=\left\{\begin{array}{ccc}
    -1&{\rm for}&x<0\\
    0&{\rm for}&x=0\\
    1&{\rm for}&x>0.\end{array}\right.\]
If $v$ is a real potential, $\Re[v]=v$ and $\Im[v]=0$. In this
case (\ref{k-for-delta}) together with (\ref{int-eq}) and
$\eta(x-y)=\delta(x-y)$ yield (\ref{u-for-L2}).

For a purely imaginary potential, $\Re[v]=0$, $v=i\Im[v]$, and
(\ref{k-for-delta}) takes the following form.
    \be
    {\cal K}\delta(x-y)=\frac{m}{\hbar^2}~{\rm sign}(y-x)
    \int^{\frac{x+y}{2}}dr~v(r).
    \label{k-for-delta-imaginary}
    \ee

\section{Applications}

\subsection{${\cal PT}$-Symmetric Square Well}

${\cal PT}$-symmetric square well potential,
    \be
    v(x):=\left\{\begin{array}{ccc}
    -i\zeta~{\rm sign}(x) &{\rm for}& |x|<\frac{L}{2}\\
    \infty &{\rm for}& |x|>\frac{L}{2},\end{array}\right.
    \label{sw}
    \ee
with $\zeta\in\R$ and $L\in\R^+$, defines one of the best-known
exactly solvable toy models that captures the generic properties
of pseudo-Hermitian quantum systems \cite{znojil,bagchi-quesne}. A
thorough investigation of the physical content of this model is
conducted in \cite{jpa-2004b} where a particular perturbative
calculation of a metric operator and the corresponding physical
observables, localized states, probability density, and the
underlying classical Hamiltonian is performed. This calculation
makes use of the fact that the non-Hermiticity effects in this
model diminish for energy states with larger spectral label $N$.
More specifically it is $\zeta/N^2$ that plays the role of the
perturbation parameter.

More recently, Bender and Tan \cite{bender-tan} performed a more
conventional perturbative calculation of a metric operator taking
$\zeta$ as the perturbation parameter. This is the metric operator
$\eta_+$ that is associated with the ${\cal CPT}$-inner product
$(\cdot,\cdot)_{_{\cal CPT}}$, \cite{bbj}. That is
$(\cdot,\cdot)_{_{\cal CPT}}=\br\cdot|\eta_+ \cdot\kt$,
\cite{jmp-2003}. Expressing $\eta_+$ in its exponential form,
$\eta_+=e^{-Q}$, and noting that Bender and Tan set
$\hbar=2m=L/\pi=1$, take $\epsilon=-\zeta$ for the coupling
constant, and use ``$\varepsilon(x)$'' for `` sign$(x)$'', we can
summarize their principal result (Eq.~(11) of \cite{bender-tan})
as
    \be
    \br x|Q|y\kt=:Q(x,y)=-\frac{i\zeta}{4}[x-y+{\rm
    sign}(x-y)(|x+y|-\pi)]+{\cal O}(\zeta^3),
    \label{Q-bt}
    \ee
where ${\cal O}(\zeta^n)$ stands for terms of order $n$ and higher
in powers of $\zeta$. In particular, in view of the identity
$x-y=|x-y|\:{\rm sign}(x-y)$, we have the following expression for
the ${\cal CPT}$-metric operator $\eta_+$.
    \bea
    \eta_+(x,y)&=&\delta(x-y)+\frac{i\zeta}{4}(|x-y|+|x+y|-\pi)\;
    {\rm sign}(x-y)+{\cal O}(\zeta^2).
    \label{eta-bt}
    \eea

The perturbative calculation of the metric operator using the
method developed in the preceding section is quite
straightforward. Inserting (\ref{sw}) in
(\ref{k-for-delta-imaginary}) and performing the trivial integral
in the resulting equation, we find
    \be
    {\cal K}\delta(x-y)=\frac{im\zeta}{2\hbar^2}\;|x+y|\;
    {\rm sign}(x-y).
    \label{k-delta-sw}
    \ee
The most general metric operator $\eta$ that reduces to the
identity operator in the Hermitian limit $\zeta\to 0$ is obtained
by setting
    \be
    u(x,y)=\delta(x-y)+\zeta[w_+(x-y)+w_-(x+y)]+{\cal O}(\zeta^2)
    \label{u-first-order}
    \ee
in (\ref{formal}), where $w_\pm:[-\frac{L}{2},\frac{L}{2}]\to\C$
are arbitrary functions satisfying $w_\pm(x)^*=w_\pm(\mp x)$ and
$w_\pm(\pm L)=0$.\footnote{These conditions arise from the
Hermiticity requirement on the metric operator and its spectral
resolution (\ref{spec}).} This together with (\ref{k-delta-sw})
yield
    \be
    \eta(x,y)=\delta(x-y)+\zeta\left[w_+(x-y)+w_-(x+y)+
    \frac{im}{2\hbar^2}\,|x+y|\,{\rm sign}(x-y)\right]+{\cal
    O}(\zeta^2).
    \label{gen-eta-sw}
    \ee
Setting $\hbar=2m=L/\pi=1$ in this equation, we find that the
${\cal CPT}$-metric operator (\ref{eta-bt}) obtained by Bender and
Tan \cite{bender-tan} is a particular example of the metric
operators (\ref{gen-eta-sw}). It corresponds to the choice
$w_+(x)=\frac{i}{4}(|x|-\pi)\,{\rm sign}(x)$ and $w_-(x)=0$.

We can calculate higher order terms in the expression for the
metric operator using our iterative method. Each additional order
will involve an arbitrary pair of functions that enter the
expression for $u$ in (\ref{formal}). This calculation is not only
completely general (as it yields the most general metric
operator), but it is also much simpler to perform. This is mainly
because unlike its alternatives \cite{jpa-2004b,bender-tan} it
avoids approximating or summing complicated series.

\subsection{An Imaginary Scattering Potential}

Consider the following variant of the ${\cal PT}$-symmetric square
well potential \cite{rdm}.
    \be
    v(x):= \frac{i\zeta}{2}\,[\,{\rm sign}(x+\mbox{\small$\frac{L}{2}$})+
    {\rm sign}(x-\mbox{\small$\frac{L}{2}$})-2\,{\rm sign}(x)]=
    \left\{\begin{array}{ccc}
    -i\zeta~{\rm sign}(x) &{\rm for}& |x|<\frac{L}{2}\\
    0 &{\rm for}& |x|>\frac{L}{2},\end{array}\right.
    \label{v}
    \ee
where $\zeta\in\R$ is a coupling constant and $L\in\R^+$ is a
length scale.

In \cite{jmp-2005} we established the reality of the spectrum of
this potential and used the spectral method of \cite{p1,p2,p3} to
obtain a perturbative expression for an associated metric operator
$\eta_+$. This involved constructing an appropriate biorthonormal
system for the model and performing highly tedious calculation of
the integrals appearing in the spectral resolution of $\eta_+$.
Indeed, this calculation could only be done after expanding all
the relevant quantities in powers of $\zeta$ and restricting to
the first order terms. Although the results reported in
\cite{jmp-2005} required performing extremely lengthy calculations
partly done using Mathematica, the expression obtained for
$\eta_+(x,y):=\br x|\eta_+|y\kt$ took a surprising simple form,
namely
    \be
    \eta_+(x,y)=\delta(x-y)+\frac{im\zeta}{4\hbar^2}
    \left(2L+2|x+y|-|x+y+L|-|x+y-L|\right)~{\rm sign}(x-y)+
    {\cal O}(\zeta^2).
    \label{eta1-sca}
    \ee

Here, we wish to use the scheme developed in the preceding section
to construct the most general metric operator $\eta$ that reduces
to the identity operator in the Hermitian limit $\zeta\to 0$. In
order to do this first we insert (\ref{v}) in
(\ref{k-for-delta-imaginary}) and perform the trivial integral in
the resulting equation to obtain
    \be
    {\cal K}\delta(x-y)=\frac{im\zeta}{4\hbar^2}
    \left(|x+y+L|+|x+y-L|-2|x+y|\right)~{\rm sign}(y-x).
    \label{k-sca}
    \ee
Substituting (\ref{u-first-order}) in (\ref{formal}) and using
(\ref{k-sca}), we then find
    \bea
    \eta(x,y)&=&\delta(x-y)+\zeta[~w_+(x-y)+w_-(x+y)+\nn\\
    &&\hspace{.5cm}
    \frac{im}{4\hbar^2}
    \left(2|x+y|-|x+y+L|-|x+y-L|\right){\rm sign}(x-y)~]+
    {\cal O}(\zeta^2),
    \label{eta-pde-gen}
    \eea
where $w_\pm:\R\to\C$ are arbitrary functions satisfying
$w_\pm^*(x)=w_\pm(\pm\, x)$. Clearly, the positive-definite inner
product (\ref{eta1-sca}) constructed in \cite{jmp-2005}
corresponds to setting $w_+(x)=\frac{imL}{2\hbar^2}~{\rm sign}(x)$
and $w_-(x)=0$.

\subsection{Imaginary $\delta$-Function Potentials}

Consider the potential
    \be
    v(x)=i\zeta\delta(x-a),
    \label{1-delta}
    \ee
where $\zeta,a\in\R$.\footnote{Clearly, we can choose the origin
of the $x$-axis so that $a=0$. We retain $a$ for future use where
we consider the multi-delta-function potentials.} It is not
difficult to solve the time-independent Schr\"odinger equation for
this potential and show that $H=\frac{p^2}{2m}+v(x)$ has a real
continuous spectrum.\footnote{$H$ is not ${\cal PT}$-symmetric.
But one may attempt to use the results of \cite{jmp-2003} to
construct a generalized ${\cal PT}$-operator (an anti-linear
involution) that commutes with $H$.} This in turn suggests that
one can construct an associated pseudo-metric operator using the
spectral method of \cite{p2,p3}. This construction is similar to
the one offered in \cite{jmp-2005} for the potential (\ref{v}). An
explicit calculation of $\eta$ using this method is however quite
involved. A much simpler construction that we will describe in the
following is based on the method of Section~3.

First, we substitute (\ref{1-delta}) in (\ref{K=}) to establish
    \be
    {\cal K} F(x,y)={\cal F}(x,y)+{\cal F}(y,x)^*,~~~~
    {\cal F}(x,y):=\frac{iz}{2}\:\theta(y-a)
    \int_{x-y+a}^{x+y-a}ds~ F(s,a),
    \label{d1}
    \ee
where $z:=2m\zeta/\hbar^2$ and $F(x,y)$ is a test function. If we
choose a $z$-independent $u$, the series expansion (\ref{formal})
becomes a power series in the coupling constant $z$. For
definiteness we shall first choose $u(x,y)=\delta(x-y)$. Setting
$F(x,y)=\delta(x-y)$ in (\ref{d1}) and using the properties of the
step function (\ref{theta}), we then find
    \be
    {\cal K}u(x,y)=\frac{iz}{2}\:\theta(x+y-2a)~{\rm sign}(y-x)
    =:u_1(x,y).
    \label{K1=}
    \ee
Alternatively, we could directly use (\ref{k-for-delta-imaginary})
to obtain (\ref{K1=}).

Next, we compute ${\cal K}^2u(x,y)$ by substituting $u_1$ for $F$
in (\ref{d1}). This yields
    \be
    {\cal K}^2u(x,y)=\frac{z^2}{4}\:[\theta(x-a)+\theta(y-a)]
    [(x+y-2a)\theta(x+y-2a)-|x-y|].
    \label{K2=}
    \ee
The higher order terms in (\ref{formal}) can be similarly
calculated. Moreover, because of the simple form of (\ref{d1}) and
(\ref{K1=}), we can actually obtain an upper bound on $|{\cal
K}^\ell u(x,y)|$ and use it to find a lower bound on the radius of
the convergence of the series (\ref{formal}).

First, we recall \cite{howie} that if a function $g:\R\to\C$ is
bounded on an interval $[\alpha,\beta]$ by some $M\in\R^+$, i.e.,
$|g(r)|<M$ for all $r\in[\alpha,\beta]$, then $|\int_\alpha^\beta
dr~g(r)|\leq M(\beta-\alpha)$. Now, let $F(x,y)$ be a function
such that $|F(s,a)|$ has an upper bound $M_{_F}$ as $s$ takes
values between $|x-y|+a$ and $x+y-a$. Then according to
(\ref{d1}),
    \be
    |{\cal K} F(x,y)|\leq |z| (|x-a|+|y-a|) M_{_F}.
    \label{bound}
    \ee
In view of (\ref{K1=}), for all $x,y\in\R$, $|{\cal K}u(x,y)|\leq
|z|/2$. To obtain an upper bound on $|{\cal K}^2u(x,y)|$ we set
$F(x,y)={\cal K}u(x,y)$ in (\ref{bound}) which allows us to
identify $M_{_F}$ with $|z|/2$ and yields for all $x,y\in\R$:
    \be
    |{\cal K}^2u(x,y)|\leq \frac{z^2}{2}\:(|x-a|+|y-a|).
    \label{K2-bound}
    \ee
We can directly verify this relation using (\ref{K2=}). Repeating
the procedure that leads to (\ref{K2-bound}), we find for all
$\ell\geq 1$ and all $x,y\in\R$: $|{\cal K}^\ell u(x,y)|\leq
|z|^\ell(|x-a|+|y-a|)^{\ell-1}/2$. This in turn implies, in view
of the elementary comparison tests, that the series (\ref{formal})
converges (absolutely) for $|z|(|x-a|+|y-a|)<1$. Hence, it
converges in an open disc in the $x$-$y$ plane that is centered at
$(x=a,y=a)$ and has a radius $\varrho>(\sqrt 2\;|z|)^{-1}$.

In summary, for every given value of $z$, truncation of the series
(\ref{formal}) yields a reliable approximation for $\eta$ provided
that we keep sufficiently large number of terms in the series and
deal with wave functions $\psi(x)$ that decay sufficiently rapidly
as $|x|\to\infty$.\footnote{Our analysis only yields a lower bound
on $\varrho$. It does not imply that $\varrho$ is finite.}

We can extend our treatment to a potential consisting of more than
one delta-function:
    \be
    v(x)=i\sum_{n=1}^N \zeta_n\delta(x-a_n),
    \label{n-delta}
    \ee
where $N\in\Z^+$ and $\zeta_n,a_n\in\R$. An example is the ${\cal
PT}$-symmetric potentials \cite{pt-delta-potential,zn,de}
corresponding to the cases that $N$ is even and
$\zeta_{\frac{N}{2}+k}=-\zeta_k$, $a_{\frac{N}{2}+k}=-a_{k}$ for
all $k=1,2,\cdots\frac{N}{2}$.\footnote{Here we assume that
$(\zeta_n,a_n)$ are such that the Hamiltonian is pseudo-Hermitian.
This is the generic case, for the values of $(\zeta_n,a_n)$ that
render the spectrum of the Hamiltonian non-real form a
measure-zero subset of the set $\R^{2N}$ of all possible values of
$(\zeta_n,a_n)$.}

For these multi-delta-function potentials the calculation of the
first order term in $z_n:=2m\zeta_n/\hbar^2$ in the series
expansion (\ref{formal}) reduces to the case $N=1$ that we
considered above. In view of (\ref{K1=}),
    \be
    {\cal K}\delta(x,y)=\frac{i}{2}\sum_{n=1}^Nz_n\:\theta(x+y-2a_n)
    ~{\rm sign}(y-x).
    \label{n-K1=}
    \ee

The most general $\eta$ that reduces to $\eta=I$ in the Hermitian
limit $z_n\to 0$ is obtained up to second order terms in $z_n$ by
setting $u(x,y)=\delta(x-y)+\sum_{n=1}^Nz_n[w_{n+}(x-y)+
w_{n-}(x+y)]+{\cal O}(z_n^2)$ where $w_{n\pm}:\R\to\C$ are
arbitrary functions satisfying $w_{n\pm}(x)^*=w_{n\pm}(\mp\,x)$.
This together with (\ref{n-K1=}) and (\ref{formal}) yield
    \be
    \eta(x,y)=\delta(x-y)+\sum_{n=1}^Nz_n
    \left[w_{n+}(x-y)+w_{n-}(x+y)+\frac{i}{2}\,\theta(x+y-2a_n)
    ~{\rm sign}(y-x)\right]
    +{\cal O}(z_n^2).
    \label{eta-delta-gen}
    \ee

We close this section by the following general remarks. As we
observe in the study of the above toy models, the series
solution~(\ref{formal}) may be used to obtain a perturbative
expansion for the pseudo-metric operator $\eta$. In general
depending on the details of the model under study one may or may
not have access to a dimensionless perturbation parameter. Typical
examples for which this occurs are the imaginary cubic potential
and the single imaginary delta-function potential (\ref{1-delta}).
In this case, as explained in \cite{p67}, the truncation of the
perturbative expansion of $\eta(x,y)$ generally yields a reliable
result only within a sufficiently small region in the $x$-$y$
plane. Furthermore, one expects that for sufficiently small values
of the perturbation parameter (the coupling constant $\zeta$ or
$\zeta_n$ in the above examples) the perturbative corrections to a
positive-definite metric operator such as $\eta=I$ leave this
property intact.

\section{Concluding Remarks}

In this article, we have outlined a differential realization of
the pseudo-Hermiticity condition that plays a central role in
devising a unitary quantum theory based on quasi-Hermitian
Hamiltonians of the standard form. The integral kernel $\eta(x,y)$
for the corresponding pseudo-metric operators $\eta$ satisfies a
linear partial differential equation. For systems having $\R$ as
their configuration space this is nothing but a particular
variable-mass Klein-Gordon equation. We have obtained a general
series solution for this equation and demonstrated its application
in treating the ${\cal PT}$-symmetric square well potential, an
imaginary ${\cal PT}$-symmetric scattering potential, and a class
of imaginary delta-function potentials. In particular for the
former two potentials, the approach presented here is by far more
practical than the alternative approaches that use the spectral
resolution of the metric operator. Another advantage of the former
approach is that it is capable of producing the most general
pseudo-metric operator. In particular, imposing the
positive-definiteness condition (\ref{positive-def}), it yields
the general form of the metric operators.

Our method is not only practically advantageous but also
conceptually appealing. It furthers the analogy between Quantum
Mechanics and General Relativity, for the differential
pseudo-Hermiticity relation plays a similar role in Quantum
Mechanics as the Einstein's field equation does in General
Relativity.\footnote{Not to mention the curious fact that the
field theoretic extension of the pseudo-Hermiticity relation
(\ref{pde}) is a functional differential equation that has the
same structure as the Wheeler-DeWitt equation of the conventional
canonical quantum gravity.} Another valuable outcome of our method
is a concrete characterization of the arbitrariness of the metric
operator. Each choice of a metric operator defines a separate
quantum system. One can pursue the prescription used in the
so-called quasi-Hermitian quantum mechanics \cite{geyer-92} to
select an irreducible set of compatible quasi-Hermitian operators
$O_\alpha$ and fix the metric operator $\eta_+$ (up to scale)
through the requirement that $O_\alpha$ be
$\eta_+$-pseudo-Hermitian. Alternatively, one can follow the
approach of the so-called pseudo-Hermitian quantum mechanics
\cite{cjp-2003a}, choose $\eta_+$ directly, and construct the
Hilbert space and observables of the theory accordingly
\cite{jpa-2004b,jpa-2005b,jmp-2005,p67}.



\ed

Even for a polynomial potential, as simple as $v(x)=ix^3$, the
differential equation that the method of \cite{scholtz-geyer}
yields is much more complicated than
(\ref{pde-1}).\footnote{Compare Eq.~(32) of \cite{scholtz-geyer}
and Eq.~(\ref{pde-1}) above with $v=igx^3$.} As explained below,
however, this method has the advantage of producing an explicit
expression for the operator $\eta$ directly.